\documentclass[pra,twocolumn,showpacs,preprintnumbers,amsmath,amssymb,floatfix,superscriptaddress]{revtex4-1}

\usepackage{graphicx}
\usepackage{dcolumn}
\usepackage{bm,bbm}

\begin{document}


\title{
Coherence rephasing combined with spin-wave storage using chirped control pulses}%

\author{Gabor Demeter}
\email{demeter.gabor@wigner.mta.hu}
\affiliation{Wigner Research Center for Physics, Hungarian Academy of Sciences, Konkoly-Thege
Mikl\'os \'ut 29-33, H-1121 Budapest, Hungary}

\date{\today}

\begin{abstract}
Photon-echo based optical quantum memory schemes often employ intermediate steps to transform
optical coherences to spin coherences for longer storage times. We analyze a scheme that uses three
identical chirped control pulses for coherence rephasing in an inhomogeneously broadened ensemble of
three-level $\Lambda$-systems. The
pulses induce a cyclic permutation of the atomic populations in the adiabatic regime.  
Optical coherences
created by a signal pulse are stored as spin coherences at an intermediate time interval, and are
rephased for echo emission when the ensemble is returned to the initial state.
Echo emission during a possible partial rephasing when the medium is
inverted can be suppressed with an appropriate choice of control pulse wavevectors. 
We demonstrate that the scheme works in an optically dense ensemble, despite control pulse
distortions during propagation. It integrates conveniently the spin-wave storage step into
memory schemes based on a second rephasing of the atomic coherences.
\end{abstract}

\maketitle

\section{Introduction}
\label{intro}

Constructing optical quantum memories is of paramount importance for several applications in optical
quantum information processing \cite{Lvovsky2009,Tittel2010,Bussieres2013}, e.g. for building
quantum repeaters \cite{Briegel1998,Simon2007}, 
or for linear optical quantum computing \cite{Knill2001,Kok2007}.
It is thus not surprising that 
devising and building such memories, devices with the capability to store and faithfully
retrieve the quantum information contained in weak (few- or single-photon) light pulses, is
currently a very lively field of research. A wide class of potential memory schemes 
use  an inhomogeneously broadened atomic ensemble as a storage medium. 
The information carried by the amplitude and phase of the signal is mapped to atomic coherences
as it is absorbed. These coherences promptly dephase, so a coherent optical response
of the ensemble is prevented. However, if the dephased coherences can be {\em rephased} at a later
time, an echo of the signal may be emitted - optical quantum
information can be retrieved.

Numerous schemes of varying complexity have been proposed and demonstrated for rephasing the atomic
coherences of the storage medium. 
For example, Controlled Reversible Inhomogeneous Broadening (CRIB)
\cite{Kraus2006,Alexander2006,Sangouard2007,Tittel2010} and 
Gradient Echo Memory (GEM) \cite{Hetet2008,Hetet2008a}
schemes operate by broadening an initially narrow absorption line artificially with an
inhomogeneous magnetic or electric field. Reversing the field also reverses the phase evolution of
atomic coherences, so dephasing can be reversed.
Atomic Frequency Combs (AFC) \cite{deRiedmatten2008,Afzelius2009} involve preparing an
absorption feature in the form of narrow, equidistant peaks. Large
bandwidth signals are absorbed by atoms in multiple peaks, which first dephase, but later rephase
spontaneously due to the discrete nature of the frequencies. To extend storage times and
achieve on-demand retrieval, optical coherences in AFC-s can be transferred to long-lived
spin coherences between metastable ground states using strong control pulses
\cite{Afzelius2010,Minar2010}.
These schemes involve a laborious preparation of the storage medium prior to signal absorption,
but were demonstrated to function even at the single photon level 
\cite{Lauritzen2010,Gundogan2012,Zhou2012,Timoney2013}.

An earlier  proposal \cite{Moiseev2001}  to store single-photon light pulses in
inhomogeneously broadened ensembles was based on classical photon echos
\cite{Abella1966,AllenEberly}, which rely on strong control pulses for coherence rephasing.
The simplest schemes of this type were shown to suffer from two major
difficulties. The first one is noise from the inverted medium at the time of
echo emission, which is incompatible with quantum information retrieval
\cite{Ruggiero2009,Sangouard2010}.
The second one is the distortion of short, intense $\pi$-pulses - which are traditionally used
as control pulses - while propagating in the optically dense medium
\cite{Ruggiero2009,Ruggiero2010}. The first one can be remedied by silencing the primary echo that
would be emitted after the first control pulse and employing a second one to invert the medium
again. The coherences can then be rephased a second time and a secondary echo will be emitted from
an uninverted medium. One protocol, termed Revival Of Silenced Echo (ROSE) \cite{Damon2011} can be
realized by choosing the propagation direction of the control pulses such that the spatial
modulation of the rephased coherences does not fulfill the phase matching condition after
the first pulse. The second flaw can be remedied by using frequency-chirped pulses that
drive Adiabatic Passage (AP) between the atomic states as control pulses. 
With these improvements, traditional photon echos are compatible with few-photon signal
storage \cite{Bonarota2013} and are functional directly on telecom wavelengths
\cite{Dajczgewand2014}.
   
Adiabatic passage driven by chirped pulses was applied in a wide variety of fields
for decades now \cite{Vitanov2001,Kral2007}. Quite recently, AP was also employed successfully in
various
quantum memory applications, even though its use in such schemes is
a somewhat subtle affair. Contrary to most applications of AP where 
only a robust population transfer is required, in quantum memory schemes the phase that the AP
process imprints onto the atomic coherences is also very important. While manipulating the atomic 
populations, the overall phase associated with the process
must also be essentially constant across the whole ensemble.
One can show, however, that when two consecutive chirped pulses with identical 
amplitude and phase dependence are used to invert an ensemble of two-level atoms twice, 
the phases associated with each of the two AP processes cancel such that the overall
phase will be the same for all atoms \cite{Minar2010,Damon2011,Pascual-Winter2013}. 
For this reason, AP  by two chirped pulses can be used
for implementing spin-wave storage in AFC memories \cite{Minar2010},
for spin-coherence rephasing in EIT based quantum memories \cite{Mieth2012,Schraft2013},
and for optical coherence rephasing to implement the ROSE scheme in two-level atoms
\cite{Damon2011,Bonarota2013,Dajczgewand2014}. 

The first advantage of AP in these schemes is, as in almost all other
applications, that the precise parameters of the control pulses are not important,
AP is robust with respect to parameter changes.
The second one is that coherence rephasing can be realized with much smaller peak
intensities than when short $\pi$-pulses are used.
This is especially important in solid state media where the damage threshold of the
crystal must not be exceeded. 
The third advantage of AP is its ability to function in optically dense ensembles.
Population transfer and coherence rephasing induced by short $\pi$-pulses are very fragile
in an optically dense medium, because
the control pulses are strongly distorted \cite{Ruggiero2009,Ruggiero2010,Demeter2013}.
Population transfer and coherence rephasing induced by AP on the other hand is much more resistant
to pulse distortion during propagation \cite{Spano1988,Demeter2013}. 
This latter is not at all trivial, because
the two successive control pulses are distorted differently - one is absorbed by the medium, while
the other one, propagating in the inverted medium, is amplified. 

In this paper we consider the interaction between a series of chirped pulses and an 
inhomogeneously broadened, optically dense
ensemble of three-level $\Lambda$-systems. We show, that 
a chirped pulse that interacts with both optical transitions of the system can realize an
adiabatic rotation of the quantum states that results in
a cyclic permutation of the atomic populations. Using three consecutive pulses, it is possible to
regain the initial populations, and, at the same time, to rephase any optical coherences in the
ensemble created by a signal prior to the control pulses. 
During one interval between the pulses, the information stored in
optical coherences initially reside in spin coherences between the two lower levels. 
This sequence of control pulses thus integrates conveniently long time spin-wave storage into the
ROSE \cite{Damon2011} protocol using control pulses from a single source. 
We investigate various aspects of the interaction relevant to coherence rephasing for photon-echo
quantum memory applications. We consider the effect of the spectral width of the atomic ensemble
relative to the full bandwidth of the control pulses. We discuss how 
echo emission during a possible partial rephasing while the ensemble is inverted can be suppressed
by spatial phase mismatching as in the original ROSE scheme. 
Furthermore, we consider the question of control pulse propagation in the ensemble and identify the
conditions under which the present scheme can rephase coherences in an optically dense sample.
Finally we discuss some constraints that the energy level spacings of a material used for the 
realization of the scheme must fulfill and mention a specific example that does so.

\section{Permutation of atomic populations with a chirped pulse}

\label{sec_permutation}

\begin{figure}[tbh]
\includegraphics[width=0.28\textwidth]{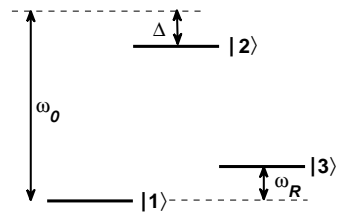}
\vspace{-1cm}
\caption{Level scheme of the atomic system. The frequency $\omega_{12}$ of the
$|1\rangle\leftrightarrow|2\rangle$ transition is offset by $\Delta$ from the inhomogeneously
broadened line center $\omega_0$. We assume that $\omega_R>0$ is the same for each atom of the
ensemble.}
\label{fig_levelscheme}       
\end{figure}

First, we study the effect of a frequency-chirped laser pulse on a single atom. 
It has three relevant energy eigenstates in a $\Lambda$-configuration (Fig. \ref{fig_levelscheme}),
the frequency of the $|1\rangle\leftrightarrow|2\rangle$ transition is
$\omega_{12}=\omega_0+\Delta$, offset by $\Delta$ from the line center $\omega_0$
of the inhomogeneously broadened ensemble. 
We assume that there is no broadening with respect to $\omega_R$ and that decoherence effects can
be neglected. 
The atomic Hamiltonian in a frame rotating with $\omega_0$ becomes
$\hat{H}_a=\hbar\Delta|2\rangle\langle2|+\hbar\omega_R|3\rangle\langle3|$.
The pulse, polarized along $\vec{e}$ interacts with both dipole allowed transitions, 
$d_{12}=\langle1|\vec{d}\vec{e}|2\rangle, d_{32}=\langle3|\vec{d}\vec{e}|2\rangle\in\mathbb{R}$
- the matrix elements are taken to be real, but not necessarily equal.
We describe the atomic state with three probability amplitudes as
$|\psi\rangle=a|1\rangle+b|2\rangle+c|3\rangle$, and, 
using the usual dipole interaction Hamiltonian and the rotating wave approximation,
seek to derive the time evolution operator
that propagates them from $t=t_1-T'$ just before
the pulse to $t=t_1+T'$ just after.

Writing $\varepsilon(t,\vec{r})=E(t,\vec{r})e^{-i\omega_0t+i\vec{k}\vec{r}}$, we can (locally)
decompose the slowly
varying complex field envelope of the pulse at the atom's location $E(t,\vec{r})$ into a
real amplitude and phase as
$\mathcal{A}(t)e^{-i\Phi(t)}=d_{12}E(t,\vec{r})/\hbar$.
Transforming to $a_r(t)=a(t),b_r(t)=b(t)e^{i\Phi(t)},c_r(t)=c(t)$, and neglecting any decoherence,
the relevant equation of motion will be:
\begin{equation}
 \partial_t
\left(\begin{array}{c}
        a_r\\ b_r\\ c_r
       \end{array}\right)
=\frac{i}{2}
\left(\begin{array}{ccc}
        0 & \mathcal{A} & 0 \\ \mathcal{A} & 2\delta & \mathcal{DA} \\ 
        0 & \mathcal{DA} & -2\omega_R
       \end{array}\right)
\left(\begin{array}{c}
        a_r\\ b_r\\ c_r
       \end{array}\right),
\label{eq_sch1}
\end{equation}
where we have introduced $\mathcal{D}=d_{32}/d_{12}$ and $\delta(t)=\partial_t\Phi(t)-\Delta$, the
instantaneous detuning perceived by
the atom. If $\mathcal{A}(t)$ and $\delta(t)$ change
sufficiently slowly, this equation can be solved by 
transforming to the reference frame of the adiabatic eigenstates at $t_1-T'$, neglecting any
nonadiabatic transitions between these eigenstates and transforming back at $t_1+T'$.

The eigenvalue problem of Eq. \ref{eq_sch1} has been discussed at length in numerous papers, most
notably in the context of Stimulated Raman Adiabatic Passage
(STIRAP) \cite{Oreg1984,Oreg1985,Gaubatz1990}. 
Here we will simply summarize the relevant results, detailed
formulas can be found for example in \cite{Fewell1997}.
First, without losing generality we assume that $\omega_R>0$. Then   
the matrix on the righthand side has three distinct ordered eigenvalues for any $\delta$, which we
denote as $\lambda^-<\lambda^0<\lambda^+$. 
In the limit when $\mathcal{A}\rightarrow 0$, the eigenvalues and the corresponding eigenvectors
$\underline{u}^{\pm,0}$ become:
\begin{align*}
\text{for} && \delta &< -\omega_R & \lambda^+&\rightarrow 0, & \underline{u}^+&\rightarrow(1,0,0)\\
&& && \lambda^0&\rightarrow -\omega_R, & \underline{u}^0&\rightarrow(0,0,1)\\
&& && \lambda^-&\rightarrow \delta, & \underline{u}^-&\rightarrow(0,1,0)\\
\text{whereas for}&& \delta &>0 & \lambda^+&\rightarrow \delta, &
\underline{u}^+&\rightarrow(0,1,0)\\
&& && \lambda^0&\rightarrow 0, & \underline{u}^0&\rightarrow(1,0,0)\\
&& && \lambda^-&\rightarrow -\omega_R, & \underline{u}^-&\rightarrow(0,0,1).\\
 \end{align*}
If nonadiabatic transitions can be neglected, the time evolution operator from $t_1-T'$ to $t_1+T'$
in the adiabatic reference frame is simply
\begin{align*}
 \hat{U}_{ad}&=\left(
\begin{array}{ccc}
 e^{i\Lambda^+} & 0 & 0\\
0 & e^{i\Lambda^0} & 0 \\
0 & 0 & e^{i\Lambda^-} \\
\end{array}\right),\\
\text{where~~}\Lambda^{\pm,0}&=\int_{t_1-T'}^{t_1+T'}\lambda^{\pm,0}(t)dt
\end{align*}
is the integral of the adiabatic eigenvalues.
For a pulse that
is chirped from blue to red such that $\delta(t_1-T')>0$, $\delta(t_1+T')<-\omega_R$,
(i.e. it becomes resonant with both atomic transitions,) the time evolution
operator in the original reference frame will be
\begin{equation}
 \hat{U}=\left(
\begin{array}{ccc}
0 & e^{i\Lambda^++i\Phi(t_1-T')} & 0 \\
0 & 0 & e^{i\Lambda^--i\Phi(t_1+T')} \\
e^{i\Lambda^0} & 0 & 0 \\
\end{array}\right).
\label{eq_Uminus}
\end{equation}
When the pulse is chirped from red to blue such that
$\delta(t_1-T')<-\omega_R$, $\delta(t_1+T')>0$, it becomes 
\begin{equation}
 \hat{U}=\left(
\begin{array}{ccc}
0 & 0 &  e^{i\Lambda^0}\\
e^{i\Lambda^+-i\Phi(t_1+T')} & 0 & 0 \\
0 & e^{i\Lambda^-+i\Phi(t_1-T')} & 0 \\
\end{array}\right).
\label{eq_Uplus}
\end{equation}

\begin{figure}[h]
\includegraphics[width=0.5\textwidth]{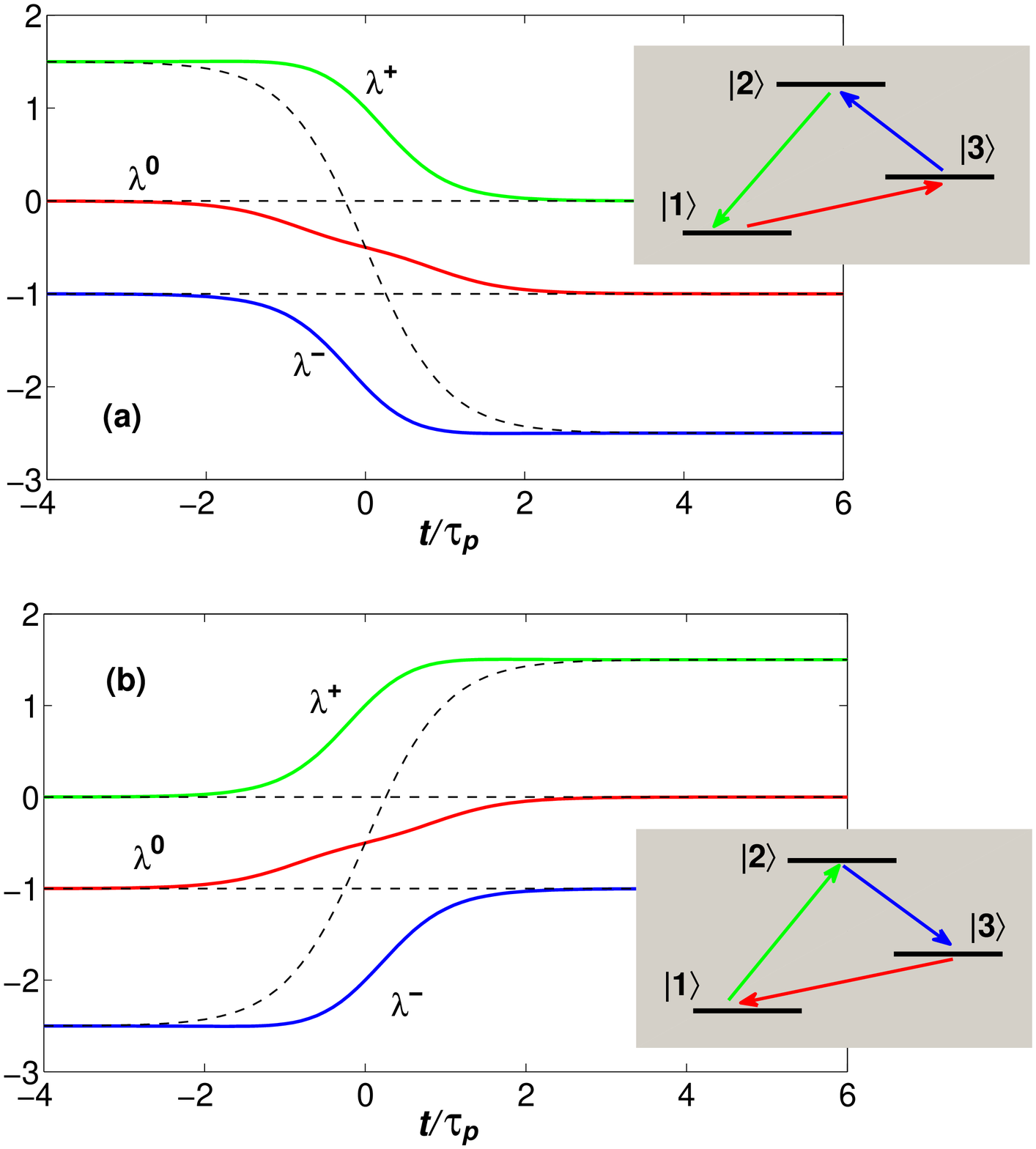}
\caption{Adiabatic permutation of atomic populations by a chirped pulse.
(a) solid lines: time evolution of the adiabatic eigenvalues for blue-to-red chirp, broken lines:
eigenvalues for $\mathcal{A}=0$ (diabatic lines). The direction of population transfer is indicated
in the inset, where each arrow is the same color as the line of the eigenvalue belonging to the
eigenstate that realizes the population transfer.
(b) eigenvalue evolution and direction of population transfer for red-to-blue chirp.
The unit of frequency in the figure is $\omega_R$, the pulse parameters are 
$\tau_p=10\omega_R^{-1}$, $A_0=2\omega_R$, $\delta_0=-0.5\omega_R$
$\mu\tau_p^{-1}=\mp2\omega_R$, and $\Delta=0$, $\mathcal{D}=1$.}
\label{fig_eigenvalues}       
\end{figure}

The matrices in Eqs. \ref{eq_Uminus} and \ref{eq_Uplus} describe a cyclic permutation
of the atomic populations with some additional phase factors. Pulses with opposite chirps 
permute the populations in an opposite sense. Figure \ref{fig_eigenvalues} illustrates
the process for a sech pulse with tanh chirp:
\begin{align}
 \mathcal{A}(t)&=A_0\text{sech}(t/\tau_p)\nonumber\\
\partial_t\Phi(t)&=\delta_0+\mu\tau_p^{-1}\tanh(t/\tau_p).
\label{eq_sech}
\end{align}
Figure
\ref{fig_eigenvalues} (a) depicts the time evolution of the adiabatic eigenvalues for 
$\mu<0$ (blue-to-red chirp), with the direction of
population transfer between the three atomic states being illustrated in the inset with gray
shading. The arrows that indicate the population transfer have been colored the same as the lines
of the corresponding eigenvalues. Figure \ref{fig_eigenvalues} (b) is a similar plot for $\mu>0$
(red-to-blue chirp).
Because we have a single field and $\omega_R\neq0$, we never have two-photon resonance during the
process - the pulse becomes resonant with the two single-photon transitions at different times.
This means that we are in a regime distinctly different from that of STIRAP - we have no dark state.
One can think of the process as two sequential adiabatic population transfers:
first on the $|1\rangle\leftrightarrow|2\rangle$, then on the
$|2\rangle\leftrightarrow|3\rangle$ transition (for blue-to-red chirp). 
On the other hand, we do have an eigenstate that starts from
$|1\rangle$ and ends in $|3\rangle$ (or the other way around).
In certain parameter ranges, this contains only a small fraction of the excited state $|2\rangle$,
at any given time, so it may be quasi-dark \cite{Djotyan2000}. 

\begin{figure}[htb]
\includegraphics[width=0.5\textwidth]{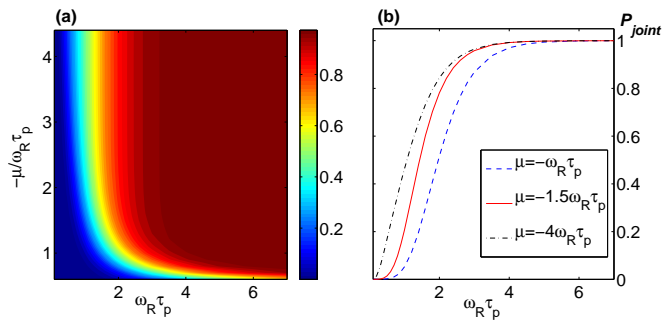}
\caption{(a) Contour plot of the joint probability that there 
is population transfer on all three atomic transitions 
$P_{joint}=|U_{12}U_{23}U_{31}|^2$ as a function of normalized pulse length $\tau_p\omega_R$ and
normalized chirp range $\mu\tau_p^{-1}/\omega_R$. (b) Line plots of $P_{joint}$ for three
values of $\mu$ as a function of $\tau_p$.
}
\label{fig_AP}       
\end{figure}

The permutation of atomic populations is robust with respect to various
parameter changes, as it is an adiabatic process.
To show this, and in particular to quantify its efficiency, we have
solved Eqs. \ref{eq_sch1} using a computer with various parameters, constructed the time evolution
matrix and calculated $P_{joint}=|U_{12}U_{23}U_{31}|^2$,
the joint probability that there is a complete population transfer on all three
transitions $|2\rangle\leftrightarrow|1\rangle$, $|1\rangle\leftrightarrow|3\rangle$ and 
$|3\rangle\leftrightarrow|2\rangle$ simultaneously. 
The pulse length $\tau_p$ and the chirp parameter $\mu$ were varied, 
while we had $\Delta=0$, $\mathcal{D}=1$ and the amplitude of the sech pulse and the
central detuning was always $A_0=20/\tau_p$ and $\delta_0=-0.5\omega_R$
(i.e. the central frequency of the pulse was exactly halfway between 
$\omega_{12}$ and $\omega_{32}$).
The results are shown in Fig. \ref{fig_AP}, where (a) shows a contour plot of $P_{joint}$ as a
function of $\tau_p$ and the chirp range $\mu\tau_p^{-1}$, normalized by $\omega_R^{-1}$ and
$\omega_R$
respectively. The line plots in (b) show $P_{joint}$  for three specific values of $\mu$ as a
function of $\tau_p$. 
It is clear from these figures that for the adiabatic 
permutation of populations to succeed, the transform limited bandwidth $\tau_p^{-1}$ must be
much less than the spacing of the two lower levels i.e. $\tau_p\omega_R\gg1$.
This condition ensures that the
two optical transitions are traversed sequentially. Figure \ref{fig_AP} (b) shows that the joint
probability is practically one for about $\tau_p\omega_R=5$. 

\section{Coherence rephasing with a series of chirped pulses}

\label{sec_rephasing}
\subsection{Control pulses with negative chirp}

\begin{figure}[htb]
\includegraphics[width=0.46\textwidth]{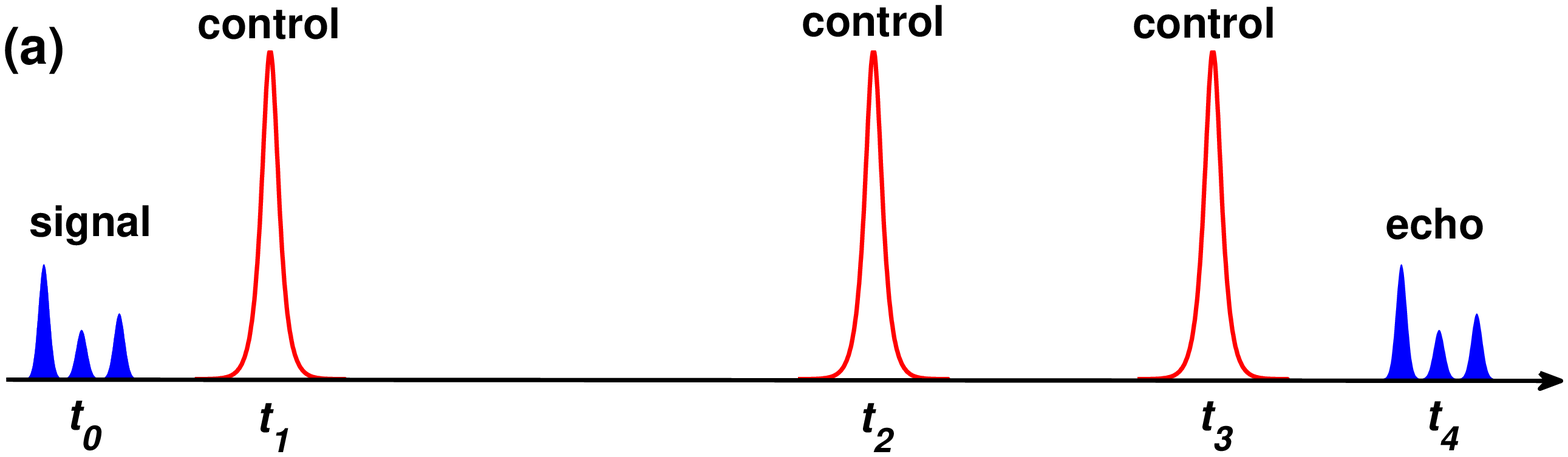}
\vspace*{0.5cm}

\includegraphics[width=0.46\textwidth]{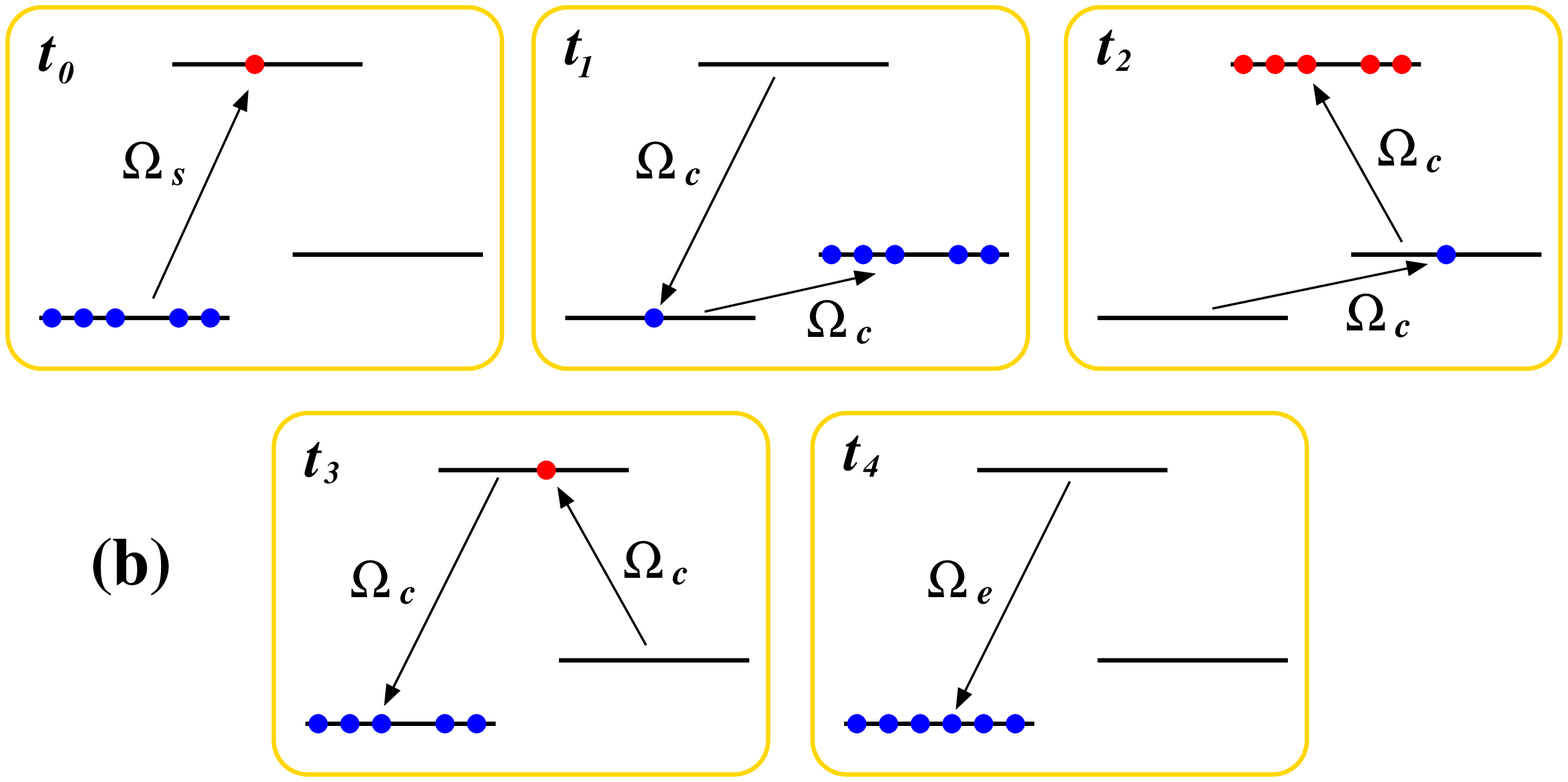}
\caption{(a) Timeline of the pulse sequence. A signal pulse at $t_0$ is followed by three control
pulses at $t_1$, $t_2$ and $t_3$. An echo of the signal is emitted at $t_4$. (b) A
sketch of the population transfers driven by the signal $\Omega_s$
at $t_0$, the three control pulses $\Omega_c$ at $t_1$, $t_2$ and $t_3$, and the echo $\Omega_e$ at
$t_4$. The figures in the rectangles symbolize the atomic states at the end of each pulse.
Optical coherences created by the signal at $t_0$ reside in spin coherences between $t_1$ and
$t_2$.}
\label{fig_timeline}       
\end{figure}

We now consider a sequence of three consecutive chirped control pulses for coherence
rephasing in an inhomogeneously broadened atomic ensemble. 
We have photon-echo based quantum memories in mind, the 
timeline of the envisioned process is sketched in Fig. \ref{fig_timeline} (a),
while the transformation of the atomic states by the various pulses is sketched in Fig.
\ref{fig_timeline} (b).
All atoms are in $|1\rangle$ initially, when 
a weak signal pulse, resonant with the $|1\rangle\leftrightarrow|2\rangle$ optical transition,
is absorbed at $t_0$. 
The $a_j^*b_j$ atomic coherences created by the signal dephase, so the overall ensemble
polarization disappears.
Three strong control pulses with negative (blue-to-red) chirp follow at $t_1$, $t_2$ and $t_3$
respectively. 
The first one at $t_1$ transforms the $a_j^*b_j$ coherences into $a_j^*c_j$ spin coherences 
between $|1\rangle$ and $|3\rangle$, where they can remain intact for  
a duration close to the spin-coherence time $T_{storage}\lesssim T_{spin}$. 
The second control pulse at $t_2$ transforms them into $c_j^*b_j$ optical coherences on the
$|3\rangle\leftrightarrow|2\rangle$ transition.
There is population inversion in the ensemble at this point, so if rephasing should
occur, echo emission must be suppressed as it will be too noisy for quantum memory applications
\cite{Ruggiero2009}. Finally, the third control
pulse at $t_3$ transforms coherences back to the $|1\rangle\leftrightarrow|2\rangle$ transition.
If coherence rephasing succeeds, there will be a revival of the ensemble polarization 
and echo emission becomes possible at $t_4$. 

To investigate whether coherence rephasing by the control pulses is indeed possible, we
construct the time evolution operator that evolves the atomic states from 
$t=t_0+T$ just after the signal pulse has been absorbed, to $t=t_4-T$ just before the 
echo is emitted at $t_4$:
\begin{multline}
\hat{U}(\Delta,\vec{r})=\hat{U}^F(t_4-T,t_3+T')\hat{U}_3(\Delta,\vec{r})
\hat{U}^F(t_3-T',t_2+T')\\
\times\hat{U}_2(\Delta,\vec{r})\hat{U}^F(t_2-T',t_1+T')
\hat{U}_1(\Delta,\vec{r})\hat{U}^F(t_1-T',t_0+T)
\label{eq_Uoperator}
\end{multline}
Here $\hat{U}_i(\Delta,\vec{r})$ is the time evolution operator of the $i$-th control pulse from
$t=t_i-T'$ to $t=t_i+T'$, while 
$\hat{U}^F(t',t)$ is that for free evolution between the control pulses:
\begin{equation}
 \hat{U}^F(t',t)=\left(
\begin{array}{ccc}
 1 & 0 & 0 \\
0 & e^{-i\Delta(t'-t)} & 0 \\
0 & 0 & e^{-i\omega_R(t'-t)} \\
\end{array}\right).
\label{eq_Ufree}
\end{equation}
The $\hat{U}_i(\Delta,\vec{r})$ depend on $\Delta$ and $\vec{r}$, because the adiabatic eigenvalues
$\lambda^{\pm,0}$ contain an explicit dependence on $\Delta$, and an implicit dependence
on $\vec{r}$ through $\mathcal{A}$ and $\Phi$ which vary in space as well. 

For atoms with $\Delta$ such that all three control pulses
permute the populations as described by Eq. \ref{eq_Uminus}, 
$\hat{U}(\Delta,\vec{r})$ will be diagonal with:
\begin{align}
\left[\hat{U}(\Delta,\vec{r})\right]_{11}&=e^{i\Lambda_1^0+i\Lambda_2^-+i\Lambda_3^+}
\times\nonumber\\
&e^{-i\Delta(t_3-t_2-2T')- i\omega_R(t_2-t_1-2T')},\nonumber \\
\left[\hat{U}(\Delta,\vec{r})\right]_{22}&=e^{i\Lambda_1^++i\Lambda_2^0+i\Lambda_3^-}
\times\nonumber\\
&e^{-i\Delta(t_1-t_0+t_4-t_3-2T-2T')- i\omega_R(t_3-t_2-2T')},\nonumber \\
\left[\hat{U}(\Delta,\vec{r})\right]_{33}&=e^{i\Lambda_1^-+i\Lambda_2^++i\Lambda_3^0}
\times\nonumber\\
&e^{-i\Delta(t_2-t_1-2T')- i\omega_R(t_1-t_0+t_4-t_3-2T-2T')}. 
\label{eq_Uoveral}
\end{align}
For brevity, we have not indicated the
$\Delta$ and $\vec{r}$ dependence of the $\Lambda_i^{\pm,0}$ and have
dropped the unimportant $\Phi_i(t_i\pm T')$ constant phase terms.
The $a^*b$ coherence at $t=t_4-T$ is then given in terms of its value at $t=t_0+T$ as:
\begin{multline}
 (a^*b)'=a^*b\exp\left(i[\Lambda_2^0-\Lambda_1^0+\Lambda_1^+-\Lambda_3^+
+\Lambda_3^--\Lambda_2^-]\right)\\
\times\exp\left(-i\Delta[t_1-t_0+t_4-2t_3+t_2-2T]\right)\\
\times\exp\left(-i\omega_R[t_1-2t_2+t_3]\right).
\label{eq_rephas1}
\end{multline}
If the three control pulses are identical, the corresponding $\Lambda_j^{\pm,0}$-s are equal,
 so the argument of the first exponential vanishes for any $\Delta$. The third exponential
describes a uniform phase shift for the whole ensemble, while the second one 
is $\exp(i2\Delta T)$ provided that the time intervals between the pulses fulfill
\begin{equation}
 t_1-t_0+t_4-t_3=t_3-t_2.
\label{eq_rephas2}
\end{equation}
Therefore with three identical control pulses, 
the atomic coherences will be rephased at $t_4$ - an echo of the signal can be emitted.
Equation \ref{eq_rephas2} also shows that rephasing does not depend on
the duration $t_2-t_1$ when the coherences generated by the signal reside
in the $|1\rangle\leftrightarrow|3\rangle$ spin coherence.
Transferring optical coherences to spin coherences for long time storage is a step in several
optical quantum memory protocols \cite{Moiseev2001,Minar2010}, and is usually executed with a
control pulse
pair used solely for this purpose. However, it is inherently and conveniently included in the
present
protocol, using control pulses from a single source.

The rephasing described above is in fact the secondary rephasing which happens around the ground
state like in the ROSE protocol. 
To show this, we consider the fate of optical coherences at a time $t_3'\in[t_2+T',t_3-T']$ 
when there is population inversion in the ensemble.
The time evolution operator from $t=t_0+T$ to $t=t_3'-T$ for atoms whose populations are permuted by
the first two control pulses will have the following nonzero elements:
\begin{align}
\left[\hat{U}'(\Delta,\vec{r})\right]_{13}&=e^{i\Lambda_1^-+i\Lambda_2^+}\times\nonumber\\
&e^{-i\Delta(t_2-t_1-2T')- i\omega_R(t_1-t_0-T-T')}, \nonumber\\
\left[\hat{U}'(\Delta,\vec{r})\right]_{21}&=e^{i\Lambda_1^0+i\Lambda_2^-}\times\nonumber\\
&e^{-i\Delta(t_3'-t_2-T-T')- i\omega_R(t_2-t_1-2T')}, \nonumber\\
\left[\hat{U}'(\Delta,\vec{r})\right]_{32}&=e^{i\Lambda_1^++i\Lambda_2^0}\times\nonumber\\
&e^{-i\Delta(t_1-t_0-T-T')- i\omega_R(t_3'-t_2-T-T')}. \nonumber
\end{align}
The $c^*b$ coherence
at $t_3'-T$ will thus be given by:
\begin{multline}
 (c^*b)'=ab^*\exp\left(i[\Lambda_1^0-\Lambda_2^0+\Lambda_2^--\Lambda_1^+
]\right)\\
\times\exp\left(i\Delta[t_1-t_0+t_2-t_3']\right)\\
\times\exp\left(i\omega_R[t_1-2t_2+t_3'-T+T']\right).\nonumber
\end{multline}
If the two control pulses are identical and $t_1-t_0=t_3'-t_2$ (which is bound to happen,
since we need $t_1-t_0<t_3-t_2$ to obtain the secondary rephasing), the only
$\Delta$ dependent term in the coherence will be $\exp(i\Lambda_2^--i\Lambda_1^+)$. This
in general will prevent a perfect rephasing, but may not prevent a partial revival 
of the ensemble polarization. There may then be a partial echo emission with the
corresponding loss of the stored information.

\subsection{Control pulses with positive chirp}

The properties of coherence rephasing with three positively
chirped pulses can be derived in an analogous manner from the time evolution matrices
\ref{eq_Uplus} and \ref{eq_Ufree}. The order in which the quantum states are traversed is now
reversed compared to that shown on Fig. \ref{fig_timeline} (b). The medium will be inverted
after the first control pulse between $t_1$ and $t_2$, while spin-wave storage
will take place between $t_2$ and $t_3$.
The relevant formula for the evolution of the atomic coherences
from $t=t_0+T$ to $t=t_4-T$ is:
\begin{multline}
 (a^*b)'=a^*b\exp\left(i[\Lambda_2^0-\Lambda_3^0+\Lambda_3^+-\Lambda_1^+
+\Lambda_1^--\Lambda_2^-]\right)\\
\times\exp\left(-i\Delta[2t_1-t_0+t_4-t_3-t_2-2T]\right)\\
\times\exp\left(i\omega_R[t_1-2t_2+t_3]\right).\nonumber
\end{multline}
Again, for three identical control pulses the argument of the first exponential vanishes,
while the second exponential gives $t_1-t_0+t_4-t_3=t_2-t_1$ for the condition of
rephasing. As before, we must also consider a possible rephasing in the inverted medium, this time
after the
first control pulse.  The $|3\rangle\leftrightarrow|2\rangle$ coherence at $t_2'-T$, will be: 
\begin{multline}
 (c^*b)'=ab^*\exp\left(i[\Lambda_1^+-\Lambda_1^-]\right)\times\\
\exp\left(i\Delta[t_1-t_0+t_1-t_2']\right)
\exp\left(-i\omega_R[t_1-t_2+T+T']\right).\nonumber
\end{multline}
The second exponential will certainly disappear at some $t_2'\in[t_1,t_2]$ because
we need $t_1-t_0<t_2-t_1$ to obtain the secondary rephasing. While the first exponential will not
be zero, it may again be too weakly dependent on $\Delta$ to extinguish the primary echo fully.

\section{Towards a quantum memory}

\subsection{Silencing the primary echo}

\begin{figure}
\includegraphics[width=0.46\textwidth]{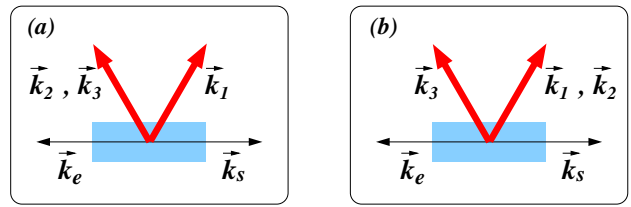}
\caption{Control pulse propagation directions for obtaining a backward echo using (a) positively
and (b) negatively chirped control pulses. The angles enclosed between neighboring wavevectors are
all $\pi/3$ in both cases. }
\label{fig_wavevectors}       
\end{figure}

As discussed for ROSE in two-level atoms \cite{Damon2011,Demeter2013}, the primary echo in
the inverted ensemble can be silenced with an appropriate choice of the control pulse wavevectors.
If the spatial modulation of the revived ensemble polarization does not fulfill the
phase matching condition, a collective emission is not possible - coherence rephasing 
does not lead to echo emission. Assuming plane wave control pulses and
taking into account the $\sim\exp(i\vec{k}_l\vec{r})$ fast spatial modulation of the $l$-th 
amplitude, we can derive the spatial modulation of the coherences after each pulse.
We can then obtain the following phase matching conditions for the echo wavevectors:
\begin{align}
\text{For negative chirp:~~~} & \vec{k}_e^{(3)}=2\vec{k}_3-\vec{k}_2-\vec{k}_1+\vec{k}_s \nonumber\\
&\vec{k}_e^{(2)}=\vec{k}_1+\vec{k}_2-\vec{k}_s\nonumber\\
\text{For positive chirp:~~~} & \vec{k}_e^{(3)}=\vec{k}_3+\vec{k}_2-2\vec{k}_1+\vec{k}_s \nonumber\\
&\vec{k}_e^{(1)}=2\vec{k}_1-\vec{k}_s
\label{eq_echos}
\end{align}
Here $\vec{k}_s$ is the wavevector of the signal field and $\vec{k}_e^{(i)}$ denotes the wavevector
of
the spatial modulation of the coherences after the $i$-th control pulse. 
We assume that the wavevectors of the control pulses and the signal field are all approximately
equal in magnitude $k_s=k_1=k_2=k_3$. (More precisely we assume that for any difference in the
wavevectors $L\Delta k\ll\pi$ where $L$ is the spatial extent of the storage medium.)
Echo emission is possible if there is a revival of the ensemble polarization due to (partial)
rephasing and $k_e^{(i)}=k_s$. The wavevector of the secondary echo we want emitted in both cases
is $\vec{k}_e^{(3)}$, whereas $\vec{k}_e^{(2)}$ and
$\vec{k}_e^{(1)}$ are the wavevectors of the primary echos we want suppressed.
From Eqs. \ref{eq_echos} we can deduce that:
\begin{itemize}
 \item[i)] For control pulses collinear with the signal, all propagating in the same direction 
$\vec{k}_1=\vec{k}_2=\vec{k}_3=\pm\vec{k}_s$, we have a forward secondary echo: 
$\vec{k}_e^{(3)}=\vec{k}_s$.
\item[ii)] If $\vec{k}_1=\vec{k}_2=\vec{k}_3=-\vec{k}_s$, i.e. we have backward propagating control
pulses, $\vec{k}_e^{(2)}=-3\vec{k}_s$ and $\vec{k}_e^{(1)}=-3\vec{k}_s$, so the primary echo is
silenced.
\item[iii)] If we are not restricted to signal and control pulse propagation along a single
direction, we can obtain a backward propagating secondary echo $\vec{k}_e^{(3)}=-\vec{k}_s$ 
with the setups sketched in Fig. \ref{fig_wavevectors}.
\end{itemize}

Note that if $\omega_R<0$ the formulas and figures for positive and negative chirps must be
interchanged.

\subsection{Ensemble spectral width and chirp range}

In section \ref{sec_permutation} and \ref{sec_rephasing} we have seen that coherence rephasing in
an ensemble requires the frequency sweep of the control pulses to encompass both optical transitions
($\omega_{12}$ and $\omega_{32}$) of the atoms to be rephased. At the same time, the distribution
of $\Delta$-s must be wide enough to absorb all frequency components of the signal -
the ensemble spectral
width $\sigma_\Delta$ must be greater than the signal bandwidth.
Let us assume, that the signal spectrum is bounded by
$\omega_0\pm\sigma_{s}$, i.e. this is the spectral range that the control pulses
must rephase. Assuming pulses as in Eqs. \ref{eq_sech}, the range of $\Delta$-s where both
optical resonances are crossed is
$\Delta\in[\delta_0-|\mu|\tau_p^{-1}+\omega_R,\delta_0+|\mu|\tau_p^{-1}]$, while the spectral
region properly rephased is expected to be slightly narrower due to a transition region of width
$\sim\tau_p^{-1}$ at the edges of this interval where nonadiabatic transitions are possible.
Centering the interval on $\Delta=0$ with $\delta_0=-\omega_R/2$, the condition for
$\tau_p$ and $\mu$ of the control pulses becomes:
\begin{equation*}
 \Delta_{max}=|\mu|\tau_p^{-1}-\omega_R/2>\sigma_{s}
\end{equation*}
Note that because we assume that $|3\rangle$ is empty initially and that the
signal field is very weak, we do not need $\omega_R>\sigma_s$ - the signal bandwidth
can be larger than the separation of the two lower states. 

After determining the minimal chirp range, we must also compare $\Delta_{max}$ to the
ensemble spectral width $\sigma_\Delta$.
This is an important question, because the unmanipulated 
absorption lines of atomic systems used in optical quantum memory experiments are often much
wider than the spectrum of the control pulses. For example, in rare-earth ion doped crystals 
that are popular candidates for such devices, very typical orders of magnitude are 
$\sim1\text{~GHz}$ for the inhomogeneous broadening, $\sim10\text{~MHz}$ for
hyperfine splittings of the lower states and $\sim1\text{~MHz}$ for signal bandwidths
\cite{Bonarota2013,Gundogan2012,Timoney2013}.
In these cases, quantum memory schemes usually work
only if the absorption line is first tailored to a sufficiently narrow range (for example, using
optical pumping). 
In the original ROSE scheme, a considerable number of atoms may be left excited 
after the second control pulse if $\pi$-pulses are employed for control and an unmanipulated
ensemble is used as a storage medium.
These are then a source of spontaneous noise during signal retrieval. 
However, if chirped pulses are used for coherence rephasing, the number
of atoms left excited in the spectrally wide ensemble can be orders of magnitude smaller  
\cite{Damon2011,Demeter2013}. 

\begin{figure}[htb]
\includegraphics[width=0.5\textwidth]{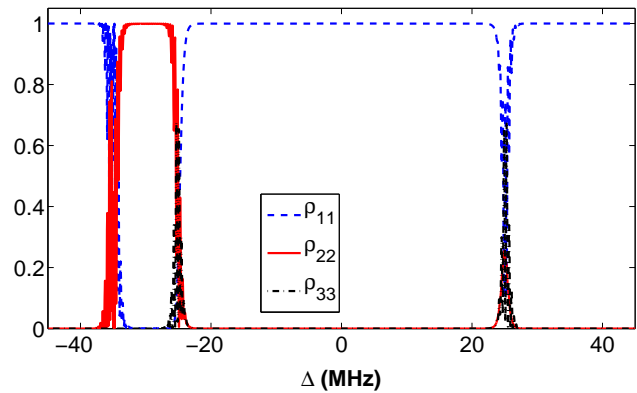}
\caption{The final atomic populations of a wide ensemble of $\Lambda$-atoms with
$\omega_R=10\text{~MHz}$ and $\mathcal{D}=1$ as a function of $\Delta$ after the three chirped
control pulses. Atoms
were in state $|1\rangle$ initially. Sech pulses with tanh chirp were used for the
calculation, with a spectrum centered on $-\omega_R/2$ and with parameters $\tau_p=1~\mu\text{s}$,
 $\mu\tau_p^{-1}=-30\text{~MHz}$ and $A_0=16 \text{~MHz}$. }
\label{fig_ensemble}       
\end{figure}

Figure \ref{fig_ensemble} depicts the effect of the control pulses on a wide ensemble.
It shows the final populations of the atomic states after the three control pulses
with parameters $\tau_p=1~\mu\text{s}$, $\omega_R=10\text{~MHz}$, $\mu\tau_p^{-1}=-30 \text{~MHz}$
and $A_0=16 \text{~MHz}$. With these parameters, atoms are rephased roughly for
$\Delta\in[-23\text{~MHz},23\text{~MHz}]$, the spectral range where
they are returned to $|1\rangle$ finally (the central part of the broken blue line on Fig.
\ref{fig_ensemble}). Apart from the transition regions where some atoms are left 
partially in $|2\rangle$ due to nonadiabatic transitions, there is also a 
wide plateau where there is full excitation after the control pulses. 
This happens because for 
atoms with $\Delta\in[-35\text{~MHz},-25\text{~MHz}]$ 
the control pulses never become resonant with $\omega_{32}$, so these atoms behave like two-level
atoms. They are inverted three times by the control pulses and end up in $|2\rangle$ after the
third one (solid red line on Fig. \ref{fig_ensemble}). So, 
in a spectrally wide ensemble, there is a $\sim\omega_R$ wide region where
atoms are left excited after the control pulses. 
This region will be a source of spontaneous noise at the time of signal retrieval. Because  
$\omega_R\gg\tau_p^{-1}$, the situation is much less
favorable than for two-level atoms, where the regions at the edge of the control
pulse bandwidth with remanent excitation were $\sim\tau_p^{-1}$ in width \cite{Demeter2013}.
Thus, to use the current rephasing scheme for quantum memory applications, 
it is necessary to tailor the absorption line to a width such that both atomic resonances are
encompassed by the control pulses.  However, this does not mean that the optical depth of the
storage medium will be reduced at the signal frequency. All atoms with resonance frequencies within
the signal spectrum are used and rephased by the control pulses.
By contrast, schemes like CRIB and GEM must first tailor an absorption line which is narrow
compared to the signal bandwidth, leading to a loss of optical depth. 

\subsection{Rephasing coherences in an optically dense ensemble}

The next question to be considered is the ability of the chirped control
pulses to rephase an optically dense ensemble. In photon-echo based memories we need optically
dense samples to absorb the signal - but these also distort the control pulses propagating within
them considerably. Moreover, consecutive control pulses will be distorted in different ways.
A pulse that excites the ensemble is absorbed, one that moves atoms back to one of the ground states
is amplified, while a pulse that moves atomic populations between the two stable states can 
propagate without considerable distortion to great optical depths \cite{Demeter2007}. Therefore even
if the
control pulses can rephase the ensemble at the entry, their ability to do so
further into the medium is a question. In previous investigations it was found, that
short $\pi$-pulses are very fragile in optically dense ensembles of two-level atoms
\cite{Ruggiero2009,Ruggiero2010}, but two consecutive chirped pulses can be used to rephase
coherences to sufficient optical depths \cite{Demeter2013}.

To answer this question, we have solved the relevant Maxwell-Bloch equation for the propagation
of the control fields using a computer. To obtain a tractable problem,  
we have assumed that the control pulses are plane waves propagating in the same direction along
the $z$ axis, i.e. we considered a one dimensional problem only. 
In the slowly varying envelope approximation, the wave equation for
the complex pulse amplitudes $\Omega_j(z,t)=E_j(z,t)d_{12}/\hbar$ will be:
\begin{multline}
 \left(\partial_z+\frac{1}{c}\partial_t\right)\Omega_j(z,t)=i\frac{\alpha_d}{\pi g(0)}\times\\
\int\left[a^*(z,t,\Delta)b(z,t,\Delta)+\mathcal{D}
c^*(z,t,\Delta)b(z,t,\Delta)\right]g(\Delta)d\Delta.
\label{eq_maxwell}
\end{multline}
Here $\alpha_d=\pi g(0)k\mathcal{N}d_{12}^2/\epsilon_0\hbar$ is the absorption constant of the
$|1\rangle\leftrightarrow|2\rangle$ transition, $g(\Delta)$ is the spectral distribution function
of the ensemble, and $\mathcal{N}$ is the density. 
The integral on the right is the overall polarization of the ensemble at time $t$ and point
$z$ - we now have an ensemble extended in both $\Delta$ and $z$, so the probability
amplitudes $a$, $b$ and $c$ depend on these variables too. 
Equation \ref{eq_maxwell} must be solved together with the optical Bloch equations
\begin{equation}
 \partial_t
\left(\begin{array}{c}
        a\\ b\\ c
       \end{array}\right)
=\frac{i}{2}
\left(\begin{array}{ccc}
        0 & \Omega_j^* & 0 \\ \Omega_j & -2\Delta & \mathcal{D}\Omega_j \\ 
        0 & \mathcal{D}\Omega_j^* & -2\omega_R
       \end{array}\right)
\left(\begin{array}{c}
        a\\ b\\ c
       \end{array}\right)
\label{eq_sch2}
\end{equation}
for the atoms of the ensemble. (This is the same as  Eq. \ref{eq_sch1} rewritten in terms of a
complex field amplitude $\Omega_j$.) The equations were solved for three consecutive,
copropagating chirped control pulses with blue-to-red chirp. Because we have optical quantum
memories in mind, the excitation that signal absorption generates in the medium is completely
negligible when considering control pulse propagation, we can assume
all atoms to be in $|1\rangle$ initially.

Having obtained $\Omega_j(t,z)$ for all three pulses, we can readily
calculate the time evolution matrices $\hat{U}_j(\Delta,z)$ by
solving Eqs. \ref{eq_sch2} again for a set of atoms with various $\Delta$-s and initial conditions.
To investigate the extent of the domain where  coherences are rephased, we define
the rephasing factor $\mathcal{R}(\Delta,z)$ using the auxiliary
quantities$\mathcal{R}_1(\Delta,z)$ and $\mathcal{R}_2(\Delta,z)$: 
\begin{align}
\mathcal{R}_1(\Delta,z)&=\left[\hat{U}_1(\Delta,z)\right]_{31}\left[\hat{U}_2(\Delta,z)\right]_{23}
\left[\hat{U}_3(\Delta,z)\right]_{12}\nonumber\\
\mathcal{R}_2(\Delta,z)&=\left[\hat{U}_1(\Delta,z)\right]_{12}\left[\hat{U}_2(\Delta,z)\right]_{31}
\left[\hat{U}_3(\Delta,z)\right]_{23}\nonumber
\end{align}
as
\begin{equation}
\mathcal{R}(\Delta,z)=\mathcal{R}^*_1(\Delta,z)\mathcal{R}_2(\Delta,z).
\label{eq_Rfactor}
\end{equation}
Comparing Eqs. \ref{eq_Uminus}, \ref{eq_Uoveral} and \ref{eq_rephas1} one can see that
$|\mathcal{R}_1(\Delta,z)|^2$ is the probability that an atom at $z$ with frequency offset
$\Delta$ in state $|1\rangle$ initially is transferred to $|3\rangle$, then to $|2\rangle$ and
finally back to $|1\rangle$ by the control pulses, while $\arg[\mathcal{R}_1(\Delta,z)]$ is the
overall phase associated with this process. $|\mathcal{R}_2(\Delta,z)|^2$ is the probability that
an atom in $|2\rangle$ initially is moved to $|1\rangle$, then to $|3\rangle$ and
finally back to $|2\rangle$, while $\arg[\mathcal{R}_2(\Delta,z)]$ is the associated phase.
For $\Delta$ and $z$ such that all three control pulses drive AP
between the atomic states,  $\mathcal{R}(\Delta,z)$ is precisely the 
first exponential factor in Eq. \ref{eq_rephas1} that contains the eigenvalue
integrals $\Lambda_j^{\pm,0}$. In general, $\mathcal{R}(\Delta,z)$ is an overall
factor whose magnitude gives the probability that all populations have been permuted three times as
required, and also the associated phase factor, whether AP has taken place or not.
Clearly, to be able to rephase the coherences in some domain of the atomic ensemble, we must have 
$|\mathcal{R}(\Delta,z)|^2=1$ and $\arg(\mathcal{R}[\Delta,z)]=\text{const.}$ in an interval of
$\Delta$ and $z$.

Note that we cannot simply construct the overall transfer matrix $\hat{U}(\Delta,z)$ and use its
diagonal matrix elements to investigate coherence rephasing.
$\hat{U}_{11}=\mathcal{R}_1$ and $\hat{U}_{22}=\mathcal{R}_2$ is true only if all three pulses
drive AP, i.e. if all elements of the $\hat{U}_j$ associated with nonadiabatic transitions are 0. 

\begin{figure}[htb]
\includegraphics[width=0.5\textwidth]{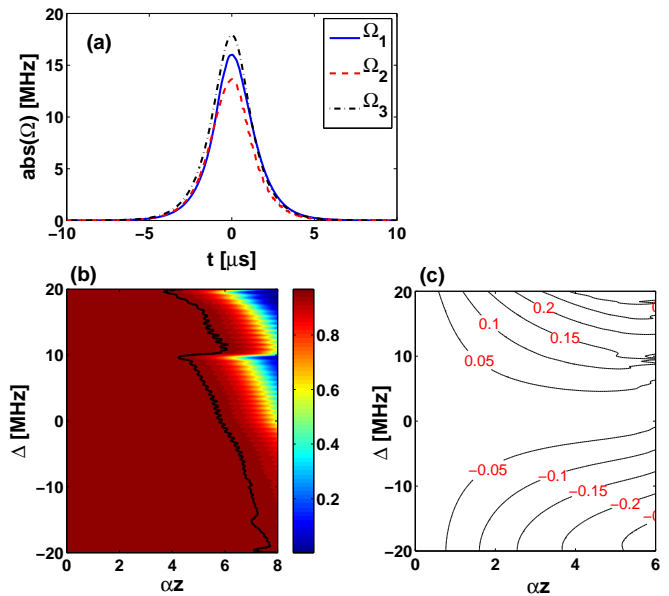}
\caption{Control pulse propagation in an optically dense ensemble. (a) Magnitude of the control
pulses at an optical depth of $\alpha_dz=5$. (b) Contour plot of 
$|\mathcal{R}(\Delta,z)|^2$ and (c) $\arg[\mathcal{R}(\Delta,z)]$ measured in radians.
$g(\Delta)=g(0)$ for $\Delta\in[-20\text{~MHz},20\text{~MHz}]$, $g(\Delta)=0$ elsewhere. Pulse
parameters are the same as for Fig. \ref{fig_ensemble}. The black line in (b) marks 
$|\mathcal{R}(\Delta,z)|^2=0.99$. }
\label{fig_propag1}       
\end{figure}

The control fields $\Omega_j$ and the rephasing factor $\mathcal{R}$ were first calculated for an
atomic ensemble whose spectral distribution was taken to be a constant in the range
$\Delta\in[-20\text{~MHz},20\text{~MHz}]$ and 0 elsewhere. 
Control pulse parameters were the same as those used 
for the calculation depicted in Fig. \ref{fig_ensemble}, so this case amounts to tailoring the
absorption line such that the pulses can perform the permutation of the atomic populations for 
whole ensemble. 
Figure \ref{fig_propag1} (a) shows the magnitude of the 
three amplitudes at an optical depth of $\alpha_dz=5$. 
Clearly, the $\Omega_j$ are different here, even though they were identical at $\alpha_dz=0$.
$\Omega_1$, which transfers atoms from $|1\rangle$ to $|3\rangle$ is not really
changed, $\Omega_2$, which excites atoms is attenuated, while $\Omega_3$, which
returns them to $|1\rangle$ is amplified. Figure \ref{fig_propag1} (b) and (c) depict
$|\mathcal{R}(\Delta,z)|^2$ and $\arg[\mathcal{R}(\Delta,z)]$ (in radians) as a
function of the optical depth $\alpha_dz$ and the frequency offset $\Delta$.
The black line in (b) marks  $|\mathcal{R}(\Delta,z)|^2=0.99$. 
The plots demonstrate that the control pulses can in fact rephase a considerable part of the
ensemble. The magnitude of $\mathcal{R}$ is very nearly 1 for the whole spectral
width until about $\alpha_dz=5$, and the phase difference is also small 
($\lesssim\pm10^{-1}\text{rad}$) at $\alpha_dz=5$ for a spectral width of about 14 MHz. 
This seems sufficient, as it would allow a signal of $\sim10$ MHz bandwidth to be stored and
emitted by the ensemble - an optical depth of $\alpha_dz=5$ allows the absorption of 99.3 \% of the
signal pulse energy. Note that for forward echos, which is the only possibility for control pulses
propagating along a single direction, $\alpha_dz=2$ is the ideal choice \cite{Sangouard2007}.

\begin{figure}[htb]
\includegraphics[width=0.5\textwidth]{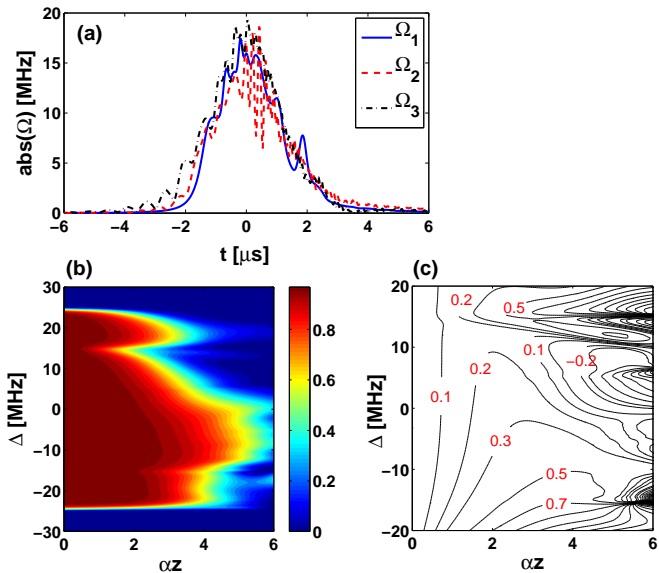}
\caption{Control pulse propagation in an optically dense ensemble. (a) Magnitude of the control
pulses at an optical depth of $\alpha_dz=5$. (b) Contour plot of 
$|\mathcal{R}(\Delta,z)|^2$ and (c) $\arg(\mathcal{R}(\Delta,z))$ measured in radians.
$g(\Delta)=g(0)$ for $\Delta\in[-50\text{~MHz},50\text{~MHz}]$, $g(\Delta)=0$ elsewhere. Pulse
parameters are the same as for Fig. \ref{fig_ensemble}.}
\label{fig_propag2}       
\end{figure}

The $\Omega_j$-s and $\mathcal{R}$ were also calculated for an atomic ensemble whose
spectral width was wide with respect to the bandwidth of the control pulses 
($g(\Delta)=\text{const.}$ was assumed in the range
$\Delta\in[-50\text{~MHz},50\text{~MHz}]$).
Figure \ref{fig_propag2} (a) depicts the $|\Omega_j|$-s at $\alpha_dz=5$, while
(b) and (c) depict $|\mathcal{R}(\Delta,z)|^2$ and $\arg[\mathcal{R}(\Delta,z)]$ (in radians).
 Clearly, this time the ability of the same control pulses to rephase the
coherences of the ensemble deteriorates much more quickly as they propagate. 
$|\mathcal{R}(\Delta,z)|^2$ drops below 0.9 by about $\alpha_dz=3$ everywhere, and the phase
differences
are also much greater than they were in the previous case. We attribute this
loss of rephasing to pulse distortions that arise from the interaction of the control pulses 
with atoms for which $\omega_{32}$ is outside the control pulse spectrum, i.e. which behave as
two-level atoms. The pulse amplitudes at $\alpha_dz=5$ shown in Fig. \ref{fig_propag2} (a)
are much more distorted than in the previous case [see Fig. \ref{fig_propag1} (a)]. 
Increasing the pulse amplitudes does not help, using 
more intensive pulses yields very similar results. Therefore we conclude that 
for the current rephasing scheme to be useful for optical quantum memory applications, the ensemble
spectral width has to be tailored to a width that allows permutation of
the atomic populations for the entire ensemble. 
However, we stress again that this does not mean that we lose any optical depth at the signal
frequency. 

\subsection{Material considerations}

Finally we consider some material properties that are required for the realization of our scheme.
A number of quantum memory schemes have been demonstrated in rare-earth (RE) doped optical 
materials, where the hyperfine levels of the dopant ions offer the possibility of selecting  
an atomic system with a $\Lambda$-configuration for spin-wave storage (see 
e.g. \cite{Afzelius2010,Gundogan2013,Timoney2013}). Apart from the various homogeneous linewidths, 
there is a special requirement in our case. Because the same control pulses are to become 
resonant with two optical transitions one after another, it is important that there be no other 
atomic transitions between the two used for realizing the $\Lambda$-linkage.
As hyperfine transitions in RE doped crystals cannot usually be polarization selected, this 
effectively means that if the excited state has multiple hyperfine sublevels, 
their separation must be greater than that of the two ground state hyperfine levels used. 

\begin{figure}[htb]
\includegraphics[width=0.46\textwidth]{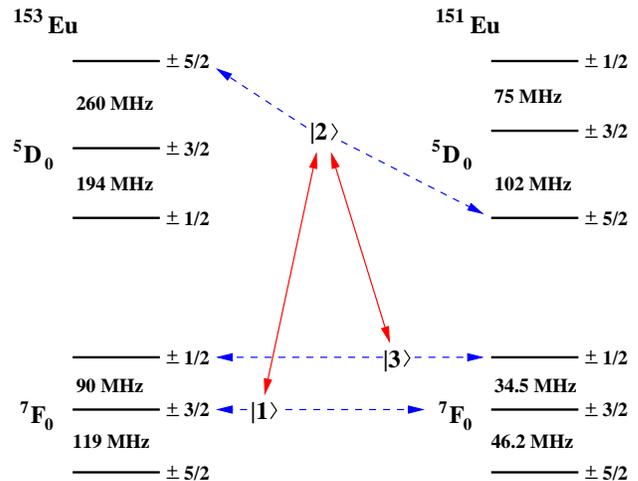}
\caption{Hyperfine energy level spacings of $\mathrm{Eu^{3+}:Y_2SiO_5}$ for the 
$^{153}\mathrm{Eu}$ isotope (from \cite{Lauritzen2012}) on the left and the $^{151}\mathrm{Eu}$ 
isotope (from \cite{Longdell2006}) on the right. The hyperfine levels used for the  
$\Lambda$-system (sketched in the middle) are connected to the relevant states with the broken blue 
arrows. Solid red arrows mark the optical transitions used.}
\label{fig_Eu}       
\end{figure}
 
One possible example of such a material is $^{153}\mathrm{Eu^{3+}:Y_2SiO_5}$, whose hyperfine 
levels have been sketched in Fig. \ref{fig_Eu} on the left, taken from \cite{Lauritzen2012}.
Making the correspondence $|1\rangle\leftrightarrow|\pm3/2g\rangle$, 
$|3\rangle\leftrightarrow|\pm1/2g\rangle$ and $|2\rangle\leftrightarrow|\pm5/2e\rangle$,
to obtain a $\Lambda$-system and denoting the frequency of the 
$|\pm3/2g\rangle\leftrightarrow|\pm5/2e\rangle$ transition by $\omega_{12}$, we have
$\omega_{32}=\omega_{12}-2\pi\times90\mathrm{~MHz}$, i.e. 
$\omega_R=2\pi\times90\mathrm{~MHz}$. There are no optical transitions between $\omega_{12}$ and 
$\omega_{32}$, the two closest to them will be the 
 $|\pm5/2g\rangle\leftrightarrow|\pm5/2e\rangle$ 
transition at $\omega_{12}+2\pi\times119\mathrm{~MHz}$ and the 
$|\pm5/2g\rangle\leftrightarrow|\pm3/2e\rangle$ 
transition at $\omega_{32}-2\pi\times51\mathrm{~MHz}$.
This means that using control pulses of 
length $\sim 0.1\mathrm{~\mu s}$ and a chirp ranging from $\omega_{12}-2\pi\times130 
\mathrm{~MHz}$ to  
$\omega_{12}+2\pi\times 40\mathrm{~MHz}$ in frequency, the angular frequency range where 
atoms can be rephased will be approximately 
$\Delta\in[-30\times2\pi\mathrm{~MHz},30\times2\pi\mathrm{~MHz}]$.
Longer control pulses (of length $\sim1\mathrm{~\mu s}$) with a slightly greater chirp range can 
rephase a range of 
$\Delta\in[-45\times2\pi\mathrm{~MHz},45\times2\pi\mathrm{~MHz}]$ 
without exciting unwanted atomic transitions. 
The two oscillator strengths are also not very different 
($\mathcal{D}\simeq2$), so driving both transitions in the adibatic regime with the same pulse is 
feasible.   
With an optical lifetime of $T_1=2\mathrm{~ms}$ and spin coherence times exceeding days for
temperatures around 2 K, \cite{Konz2003}, this systems seem fit for the realization of ROSE 
combined with spin-wave storage. 

Of course, generating control pulses with a chirp spanning this range can be quite 
challenging. A possible alternative is to use the  
$^{151}\mathrm{Eu}$ isotope, where the hyperfine level spacing is about three times 
smaller \cite{Longdell2006}, see the right side of Fig \ref{fig_Eu}. 
Then $\omega_R=2\pi\times34.5\mathrm{~MHz}$, and all resonance frequencies
are higher than the $\omega_{12}$ frequency that belongs to the 
$|\pm3/2g\rangle\leftrightarrow|\pm5/2e\rangle$ transition except for $\omega_{32}$.
We have $\omega_R=2\pi\times34.5\mathrm{~MHz}$ and the nearest unwanted resonance frequency is
 $\omega_{12}+2\pi\times46.2\mathrm{~MHz}$ belonging to the 
$|\pm5/2g\rangle\leftrightarrow|\pm5/2e\rangle$ transition. 
This system thus yields a somewhat smaller frequency range that can be used for signal storage, but 
requires a much smaller chirp range.

\section{Summary}

In this paper we have investigated coherence rephasing in an inhomogeneously broadened ensemble of
$\Lambda$-atoms with three consecutive frequency-chirped control pulses for optical quantum memory
applications. We have shown, that if the transform limited bandwidth is
much smaller than the frequency difference of the lower energy levels $\tau_p^{-1}\ll\omega_R$, but
the overall bandwidth is greater, the control pulses can drive a cyclic permutation of the 
atomic populations in the adiabatic regime. With three such pulses interacting with the ensemble
one after the other, it is possible to rephase the optical coherences left behind by 
a weak signal pulse, leading to the emission of a signal echo. We have shown that this
rephasing, which happens when the atoms are predominantly in the ground state (i.e. the medium is
not inverted), is analogous to the secondary rephasing of the ROSE scheme when two-level atoms are
being rephased with two control pulses. There may also be a partial rephasing after one or
two control pulses, when the medium is still inverted. Echo emission at this time can
be prevented by choosing the control pulse propagation directions such that the primary echo fails
the spatial phase matching condition. 
At one point during this three-pulse rephasing process, (after one or two control pulses, depending
on the sign of $\omega_R$ and the chirp direction,) the coherences left by the signal are stored as
spin coherences between the two lower levels of the $\Lambda$-system. Thus this scheme of coherence
rephasing conveniently incorporates long time spin-wave storage into the ROSE scheme using control
pulses from a single source. The separation of the lower levels $\omega_R$ limits only the control
pulse duration, but not the signal bandwidth that can be stored in the ensemble.

We have also investigated whether the current scheme is able to
rephase the coherences of an optically dense storage medium. 
We have found, that despite the fact
that the control pulses are distorted during propagation, coherence rephasing works
well in a considerable domain in terms if spectral width and optical depth. The necessary condition 
for this is, that the ensemble spectral width must be narrower than the bandwidth where 
the control pulses can rephase the atoms i.e. the control pulses must cross both optical
transitions for the entire spectral range of the ensemble.
 
We have also considered the application of the scheme in 'naturally' inhomogeneously broadened media
where the broadening is greater than the control pulse spectrum. In this case however,
there will be a $\sim\omega_R$ wide part of the ensemble where the control pulses interact with only
one of the atomic transitions. On the one hand, this part of the ensemble will remain in the
excited state after the third control pulse and will be a source of noise during signal retrieval. 
On the other, it will distort propagating control pulses much more in an
optically dense medium, so the rephasing ability of the three pulses deteriorates fast. Thus the
ensemble must be tailored to a width narrower than the control pulse spectrum before signal
absorption. However, this does not mean that optical depth is lost at the signal frequency as in
numerous other schemes.

Finally we have argued that to realize the scheme, we need materials whose excited state sublevel 
separation is greater than the ground state one and have shown that $\mathrm{Eu^{3+}}$ doped 
into $\mathrm{Y_2SiO_5}$ is a good candidate.

\bibliography{/home/gdemeter/fiz/pulseprop/publ/bibliography}

\bibliographystyle{apsrev4-1}


\end{document}